\documentstyle[epsf]{l-aa}

\begin{document}

   \thesaurus{11         
              (11.09.1;
               11.11.1;
               12.04.1)}
   \title{Decomposition of the rotation curves of distant field galaxies}

   \author{B. Fuchs$^{\rm 1}$, C. M\"ollenhoff$^{\rm 2}$,
           and J. Heidt$^{\rm 2}$}

   \offprints{B. Fuchs}

   \institute{$^{\rm 1}$Astronomisches Rechen-Institut Heidelberg,
              M\"onchhofstr. 12-14, 69120 Heidelberg, Germany\\
              $^{\rm 2}$Landessternwarte, K\"onigstuhl, 69117 Heidelberg,
Germany}

   \date{Received; accepted  }

   \maketitle
   \markboth{B. Fuchs et al.~: Decomposition of the rotation curves
            of distant field galaxies}{B. Fuchs et al.~: Decomposition
            of the rotation curves of distant field galaxies}

   \begin{abstract}

   We present decompositions of the rotation curves of distant spiral galaxies
   into contributions due to their bulges, disks, and putative dark haloes. In
   order to set constraints on the ambiguities of the decompositons we
   interpret
   the morphology of the spiral structures quantitatively in the framework of
   density wave theory. Galaxy models constrained in such a way show that the
   distant galaxies, which are much younger than nearby galaxies, are indeed
   also imbedded in dark haloes as expected from contemporary theories of the
   cosmogony of galaxies.

   \end{abstract}

%

\section{Introduction}

Vogt et al.~(1996; hereafter referred to as VFP) have determined recently
by spectroscopy with the Keck telescope optical rotation curves of a sample
of distant field galaxies. They present also quantitative photometry of the
surface brightness of the galaxies using images taken with HST. These data
allow first insight into the dynamics of distant spiral galaxies, which are
much younger than nearby galaxies. Contemporary theories of the cosmogony of
galaxies predict that these galaxies will be imbedded in dark haloes like
nearby galaxies, because the dark haloes are thought to be the sites of
galaxy formation. The data by VFP provide the first chance to search
directly for evidence of such dark haloes around distant galaxies. Quillen \&
Sarajedini (1998) have pointed out that the morphology of the disks of that
galaxies, especially their spiral structure, can be used to constrain the
dynamical properties of the disks. They concentrated on estimating the
mass -- to -- light ratios and the velocity dispersions of the galactic
disks. In this study we wish to demonstrate how the morphological
appearance of the spiral structure of the disks may be used to constrain the
otherwise ambiguous decomposition of the rotation curves of the galaxies.

Two galaxies in the sample of VFP,
VFP\,J074-2237 at a redshift of z = 0.15 and VFP\,J0305-00115 at a redshift
of
z = 0.48, have clearly evident, prominent spiral structures. There is a further
interesting galaxy, VFP\,J064-4442 at a redshift of z = 0.88, which shows
indications of spiral structure. However, the signal -- to -- noise level is
too low to resolve the structures well enough for a quantitative analysis.

In sections 2 and 3 we construct dynamical models for the two medium --
redshifted galaxies and discuss implications of the evidence for the presence
of dark matter in the galaxies in the final section.

\section{Bulge -- disk decompositions}

In the case of VFP\,J0305-00115 we adopt the decomposition of the surface
brightness distribution of the galaxy into bulge and disk components,
respectively, given by Forbes et al.~(1996). The disk is fitted by the surface
brightness profile of an exponential disk,
\begin{equation}
\Sigma_{\rm d}(R) = \Sigma_{\rm d0} \exp(-R/h)\,,
\end{equation}
with an extrapolated central surface
brightness of 20.9 mag/arcsec$^{\rm 2}$ in the I band and,
adopting a scale conversion factor of 5.25 kpc/arcsec,
a radial scale length of $h$ = 5.1 kpc. The bulge
has been fitted by Forbes et al.~(1996) by a de Vaucouleurs profile with an
effective radius of $R_{\rm e}$ = 0.8 kpc and a surface brightness of 20.5
mag/arcsec$^{\rm 2}$ in the I band at radius $R_{\rm e}$. For reasons of
easier handling of the dynamical models we
have used instead of a de Vaucouleurs law the surface brightness profile of a
bulge model with a density distribution of the form
\begin{equation}
\rho_{\rm b}(r) =
\rho_{\rm b0} \left( 1 + {\frac{r^2}{r_{\rm c,b}^2}}
\right)^{-\frac{3.5}{2}}\,.
\end{equation}
The surface brightness profile follows a similar law with the exponent lowered
by 1/2. We have fitted such a surface brightness profile to the de Vaucouleurs
law over the distance range $R$ = 0.1 to 0.6 arcsec, where the
surface
brightness of the galaxy is dominated by the bulge (Forbes et al.~1996),
and find that the de Vaucouleurs
law can be fitted with an accuracy better than 0.05 mag/arcsec$^{\rm 2}$. We
determine in this way a core radius of $r_{\rm c,b}$ = 2.8 kpc and a central
surface brightnes of 20.4 mag/arcsec$^{\rm 2}$.

The photometric parameters of VFP\,J074-2237 have been only partially published
up to now.
We have therefore retrieved its I band images from the HST archive
(prop.~id.~5109) and
have fitted the two -- dimensional surface brightness distribution of a highly
inclined (i = 80$^\circ$) exponential disk and a bulge with a surface
brightness profile
according to equation (2) to the data using the code of M\"ollenhoff (1998).
Although the fit is hampered somewhat by the very prominent spiral structure,
we find a radial disk scale length of $h$ = 3.4 kpc, if a scale conversion
factor of 2.41
kpc/arcsec is assumed, which is in excellent agreement with the value given by
VFP. The surface brightness of the bulge is low. The central surface brightness
of the disk is brighter by 3.2 mag/arcsec$^{\rm 2}$ than the central surface
brightness of the bulge. Thus we neglect the contribution by the bulge in the
following.

\section{Decomposition of the rotation curves}

\begin{figure}[htbp]
\begin{center}
   \leavevmode
      \epsffile{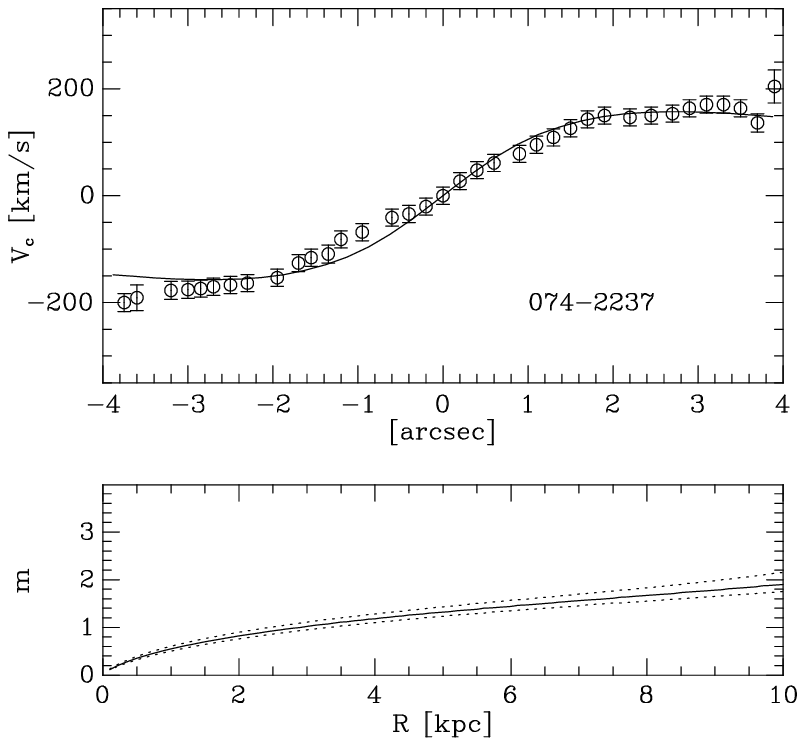}
\caption{Upper panel: Model rotation curve of VFP\,J074-2237 fitted to the
radial velocity data of VFP.
Lower panel: Number of expected spiral arms. The dotted lines indicate error
estimates.}
         \label{vfp1}
   \end{center}
   \end{figure}

Due to the comparatively large width of the slit of the spectrograph VFP have
not observed directly the rotation curves of the galaxies. But they use a
simple model of the velocity fields of the disks in which the circular velocity
is assumed to rise linearly with radius up to one radial disk scale length and
then
to remain flat. Folded with the surface brightness distribution and averaged
over the area of the slit such models can be excellently fitted to the observed
data. We adopt model rotation curves of the form
\begin{equation}
v_{\rm c}^2(R) = v_{\rm c,b}^2(R) + v_{\rm c,d}^2(R)\,,
\end{equation}
where $v_{\rm c,b}$ and $v_{\rm c,d}$ denote the contributions due to the
bulge and disk, respectively. The bulge contribution is given
according to equation (2) by
\begin{equation}
v_{\rm c,b}^2(R) = \frac{4\pi G\rho_{\rm b0}}{R} \int_0^R dr
r^2\left(1
+ {\frac{r^2}{r_{\rm c,b}^2}} \right)^{-\frac{3.5}{2}}\,,
\end{equation}
where $G$ denotes the constant of gravitation.
The rotation curve of an infinitesimally thin exponential disk is given by
\begin{equation}
v_{\rm c,d}^2(R) = 4\pi G \Sigma_{\rm d0} h x^2\left(I_0(x)K_0(x) -
I_1(x)K_1(x)\right)\, ,
\end{equation}
where $\Sigma_{\rm d0}$ denotes now the central face-on surface density of the
disk. $x$  is an abbreviation for
$x=R/2h$ and $I$ and $K$ are Bessel functions (cf. Binney \& Tremaine 1987).
The model rotation curves have been treated in the same way as by VFP. The
velocity fields and the emission-line surface brightness were projected onto
the sky adopting the inclination angles determined by VFP. As suggested by VFP,
the emission-line surface brightness profiles were approximated by exponential
disks with radial scale lengths 1.5 times that measured from the HST images.
The models were then convolved with Gaussians in order to model the blurring of
the velocity fields by seeing, which VFP estimate as 1 and 0.8 arcsec (FWHM)
in the cases of VFP\,J074-2237
and VFP\,J0305-00115, respectively. Finally the models were masked according
to the slit positions indicated in Fig.~1 of VFP. Fits of model rotation
curves obtained in this way to the observed radial velocities are shown in
Figs.~ 1 and 2 and the resulting parameters are summarized in Table 1. VFP
estimate that the uncertainties of the radial scale lengths determined from the
HST
images are about 15\%. We find that in the case of VFP\,J035-00115 the radial
velocities within 1 arcsec from the center of the galaxy can be fitted better,
if we use instead of the value given by Forbes et al.~(1996) a radial scale
length reduced by 15\% to 4.3 kpc, which we adopt for our dynamical models.

%
%
\begin{table}[t]
\caption{Dynamical parameters}
\label{Parms}
\begin{center}
\begin{tabular}{cccccc}
\hline
 VFP & $\rho_{\rm b0}$ & $\Sigma_{\rm d0}$ & $\rho_{\rm h0}$  &
 $M_{\rm lum}/{M_{\rm dark}}^{\rm a}$ & $M/L_{\rm B}$ \\[0.5ex] \hline
 074-2237 & -- & 1400 & -- & -- & 3.9\\
          & -- & 850 & 0.025 & 1.0 & 2.4\\
 0305-00115 & 0.21 & 800 & -- & -- & 4.8\\
           & 0.16 & 600 & 0.016 & 2.1 & 3.6\\
  & $M_\odot$ & $M_\odot$ & $M_\odot$ & & $M_\odot$\\
  & $pc^{-3}$ & $pc^{-2}$ & $pc^{-3}$ & & ${L_{{\rm B}\odot}}^{-1}$\\ \hline
\end{tabular}
\end{center}
\begin{list}{}{}
\item[$^{\rm a}$] Within radius $R$ = 10 kpc.
\end{list}
\end{table}

We have chosen both galaxies because of their clearly discernible spiral
structure. In order to be able to develop spiral structure galactic disks must
be dynamically cool enough, i.e.~the Toomre stability parameter $Q$
(cf.~Binney \& Tremaine 1987) must lie in the range (Quillen \& Sarajedini
1998)
\begin{equation}
 1 < Q = \frac{\kappa\sigma_{\rm U}}{3.36 G\Sigma_{\rm d}}\, \widetilde{<}\,
 2\,,
\end{equation}
where $\kappa$, $\kappa^2=2\left(\frac{v_{\rm c}}{R}\right)^2 \left(1+\frac{d
ln v_{\rm c}}{d ln R}\right)$, and $\sigma_{\rm U}$ denote the epicyclic
frequency and the radial velocity dispersion of the stars, respectively.
Furthermore, both galaxies are not grand -- design spirals. Thus their
spiral arms are almost certainly formed during `swing -- amplification'
events (Toomre 1981). This mechanism is most effective, if
the circumferential wave length of the density waves is twice the critical
wave length,
\begin{equation}
\lambda = 2 \lambda_{\rm crit} = \frac{8 \pi^2 G \Sigma_{\rm d}}{
\kappa^2}\,.
\end{equation}
The expected number of spiral arms is then
\begin{equation}
 m = \frac{2 \pi R}{\lambda}\,.
\end{equation}
The multiplicity of spiral arms has been discussed previously using similar
arguments by Athanassoula (1988), Athanassoula et al.~(1987), or Fuchs et
al.~(1996). As shown in the lower panels of Figs.~1 and 2
the predicted number of spiral arms is less than two, which is in
clear contradiction to the morphological appearance of the galaxies. Both
have bisymmetric spiral arms in the inner parts of the disks and additional
filaments in the outer parts (cf.~Fig.~1 of VFP). In the case of a single
exponential disk the amplitude of the rotation curve cancels out of
equation (7), and thus the determination of $m$. The only remaining uncertainty
is then due to the uncertainty of the radial scale length $h$. This is
illustrated for VFP\,J074-2237 in the lower panel of Fig.~1 by dotted lines.

\begin{figure}[htbp]
\begin{center}
   \leavevmode
      \epsffile{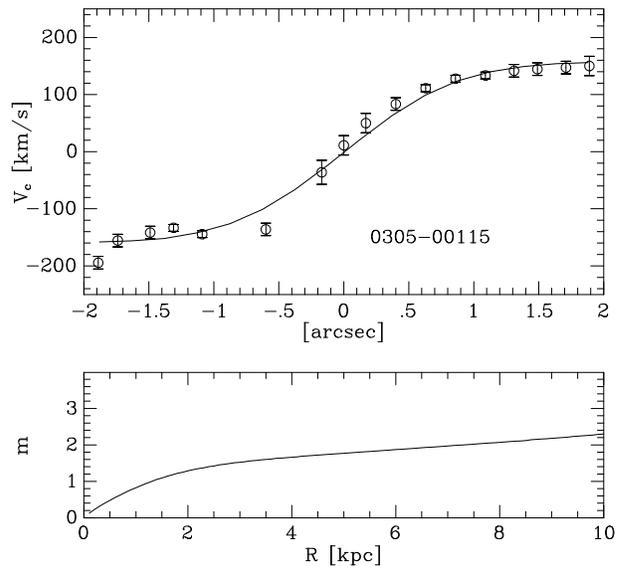}
\caption{Upper panel: Model rotation curve of VFP\,J0305-00115 fitted to the
radial velocity data of VFP. Lower panel: Number of expected spiral arms.}
\label{vfp2}
\end{center}
\end{figure}
\begin{figure}[htbp]
\begin{center}
   \leavevmode
      \epsffile{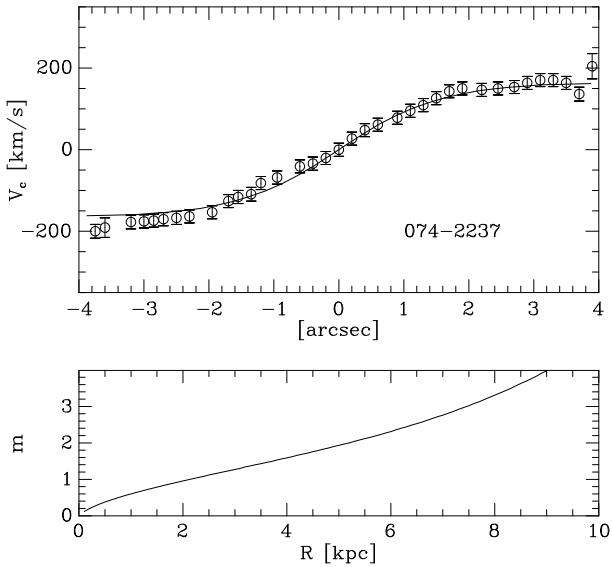}
\caption{Upper panel: `Medium' disk decomposition of the rotation curve of
VFP\,J074-2237.
Lower panel: Number of expected spiral arms.}
\label{vfp3}
\end{center}
\end{figure}
\begin{figure}[htbp]
\begin{center}
   \leavevmode
      \epsffile{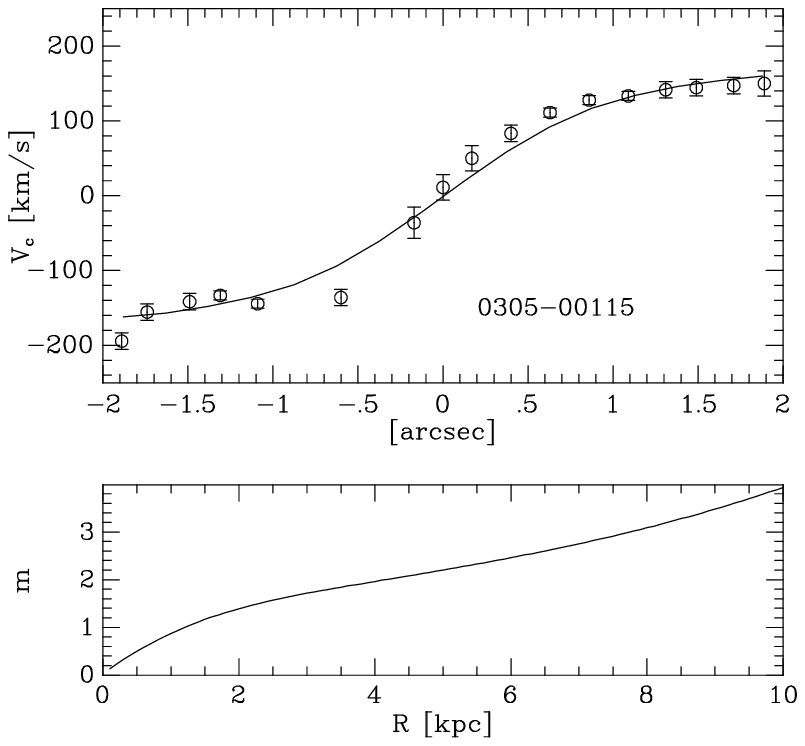}
\caption{Upper panel: `Medium' disk decomposition of the rotation curve of
VFP\,J0305-00115.
Lower panel: Number of expected spiral arms.}
\label{vfp4}
\end{center}
\end{figure}

Since it is generally expected that distant galaxies are imbedded in dark
haloes like nearby galaxies, we have also considered an additional dark halo
component in the decomposition of the rotation curves.
The dark haloes are modelled by quasi -- isothermal spheres,
\begin{equation}
\rho_{\rm h}(r) = \rho_{\rm h0} \left(1 + \frac{r^2}{r_{\rm c,h}^2}
\right)^{-1}\,,
\end{equation}
which leads to a further term,
\begin{equation}
v_{\rm c,h}^2(R) = 4\pi G \rho_{\rm h0} r_{\rm c,h}^2\left(1 -
\frac{r_{\rm c,h}}{R} arctan \frac{R}{r_{\rm c,h}}\right)\, ,
\end{equation}
in equation (3).
Unfortunately, the rotation curves span radially only two or three disk
scale lengths. As is well known from nearby galaxies, the disk and dark halo
parameters are not well constrained in such cases (van der Kruit 1995). We
have experimented with a variety of possible decompositions of the rotation
curves and find a tendency that in near `maximum' disk decompositions the
central densities of the dark haloes is low and their core radii are large
and reverse in `medium' disk decompositions. From these we have selected
models in which the expected number of spiral arms is in accordance with the
observed morphology of the galaxies. In Figs.~3 and 4 models are illustrated
with core radii of the dark haloes twice the radial disk scale lengths, which
provide fits to the observed data of the same quality as the single disk or
disk -- bulge models shown in Figs.~1 and 2. The
central densities are listed in Table 1. The disks contribute at the peaks of
the disk rotation curves 76\% and 70\% to the composite rotation curves of
VFP\,J074-2237 and VFP\,J0305-00115, respectively. Similar decompositions of
the rotation curves of nearby galaxies have been suggested by Bottema (1998).

\section{Discussion and Conclusions}

\begin{figure}[htbp]
\begin{center}
   \leavevmode
      \epsffile{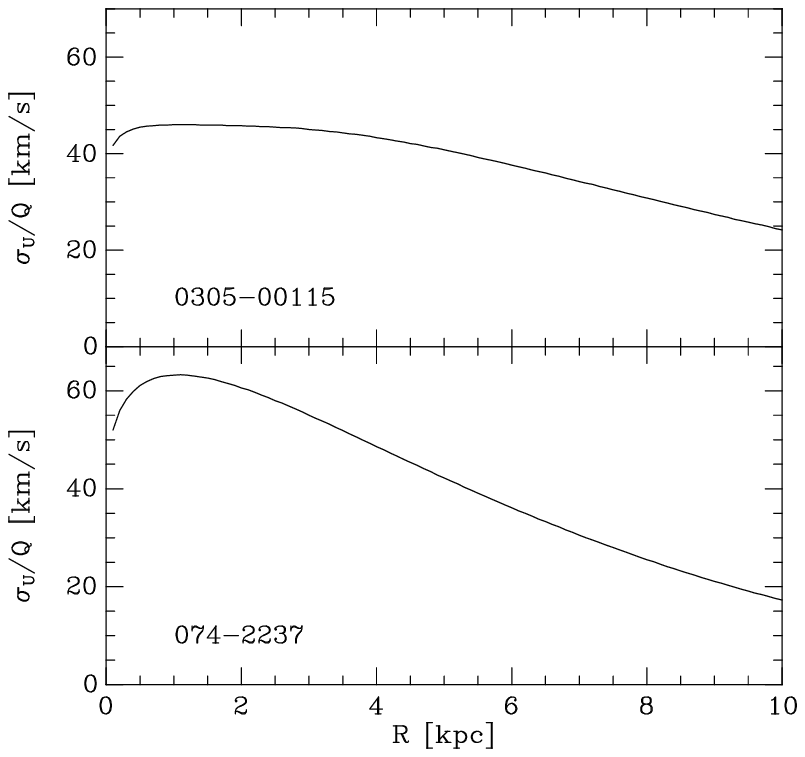}
\caption{Radial variation of the ratio of the radial velocity dipersion of
         the stars in the disks of the galaxies divided by the Toomre stability
         parameter $Q$. According to density wave theory the stability
         parameter is expected to lie in the range $1 < Q\,\widetilde{<}\,2$.}
\label{suq}
\end{center}
\end{figure}

We conclude that the interpretation of the morphology of the distant galaxies
gives circumstantial but significant evidence for the presence of dark haloes.
The galaxies are, however, in the parts observed by VFP not
dominated by dark matter as
can be seen from column (5) of Table 1. This is quite similar to those nearby
galaxies which have been studied only with optical rotation curves (Broeils
1992). Using the rest frame luminosities given by VFP we have determined the
mass -- to -- light ratios listed in the last column of Table 1. The maximum
disk mass -- to -- light ratios lie at the upper end of the range determined by
Broeils
(1992) and Broeils \& Courteau (1997) by maximum disk decompositions of
rotation curves of bright nearby spiral galaxies. Because of the dynamical
implications explained in the previous section, we argue in favour of the lower
mass -- to -- light ratios given in Table 1.

Following Quillen \& Sarajedini (1998) we discuss finally constraints set
by the Toomre stability parameter. In Fig.~5 the radial variation of the ratio
$\sigma_{\rm U}/Q$ is shown, which has been calculated according to equation
(6) for both galaxies using the `medium' disk parameters given in Table 1.
According to density wave theory the stability parameter must lie in the range
$1 < Q\,\widetilde{<}\,2$ (Toomre 1981). Thus we predict for the disk of
VFP\,J074-2237 at the radius $R = h$ a radial velocity dispersion of the stars
in the range
$\sigma_{\rm U}$ = 54 to 108 km/s, and similarly for VFP\,J0305-00115
$\sigma_{\rm U}$ = 42 to 84 km/s. We note that these values fit nicely to
velocity
dispersions observed in the disks of nearby galaxies. Bottema (1993) finds in
nearby galaxies that on the average $\sigma_{\rm U}$ = -17$\cdot{\rm M}_{\rm
B,disk}$ - 279 km/s at $R = h$,
which would imply 87 km/s for VFP\,J074-2237 and 80 km/s for
VFP\,J0305-00115, respectively, if the rest frame luminosities given by VFP
are used. However, it remains at present unclear how the local velocity
dispersion -- magnitude relation can be extended to distant galaxies. It is
generally believed that redshifted galaxies undergo some luminosity evolution.
VFP find, for instance, that the Tully -- Fisher relation defined by the
galaxies of their sample is shifted relative to the local Tully -- Fisher
relation by $\Delta M = 0.6$ mag in the sense that the redshifted galaxies are
intrinsically brighter. If the velocity dispersion -- magnitude relation is
affected in the same way, the velocity dispersions predicted from this relation
would be lower by 10 km/s, but still fully consistent with the velocity
dispersions derived from stability arguments. If the `maximum' disk
parameters are inserted into equation (6), the predicted velocity
dispersions increase by 30\% to 40\% and become unrealistically large as was
previously noted by Quillen \& Sarajedini (1998), who assumed `maximum' disks.
In the case of VFP\,J074-2237, for instance, the predicted velocity dispersion
of the disk stars would be of the order of the
expected velocity dispersion of halo stars, $v_{\rm c}/\sqrt{2}$ = 141 km/s.

Quillen \& Sarajedini (1998) have also pointed to the fact that the gaseous
disks of the galaxies cannot be too massive, because they would otherwise be
dynamically unstable, $Q_{\rm g} < 1$. Such gas disks would trigger very
violent dynamical
reactions both of the gaseous and the stellar disks (Fuchs \& von Linden 1998).
The disks would heat up on a very short time scale and the galaxies would
change their morphology from well ordered spiral structure to a highly
flocculent appearance. For VFP\,J074-2237 we find, again at the radius $R =
h$, a value of the stability parameter of the gas disk of $Q_{\rm g}(R=h) =
(\sigma_{\rm g}/54{\rm km/s})/f$, where $f$ denotes the fraction of the total
surface density of the disk in the form of interstellar gas. If we assume a
velocity dispersion of the gas of $\sigma_{\rm g} = 6$ km/s, the gas fraction
is restricted to values $f < 11\%$, although this may
increase in the outer parts of the disk. We find
for VFP\,J0305-00115 a similar low value of the gas fraction of $f < 14\%$ at
the radius $R = h$.

We conclude that, taken all arguments together, distant spiral galaxies are
indeed imbedded in dark haloes similar to those of nearby galaxies.
%
%
\begin{acknowledgements}
We gratefully acknowledge valuable comments on the manuscript of this paper by
the referee A.~Bosma.
J H is supported by the Sonderforschungsbereich 328 ``Entwicklung von
Galaxien''.
\end{acknowledgements}

\end{document}